\documentclass[showpacs,preprintnumbers,amsmath,amssymb]{revtex4}

\usepackage{epsfig}
\usepackage[colorlinks,linkcolor=blue,anchorcolor=blue,
            citecolor=blue,urlcolor=blue,
            breaklinks=true]{hyperref}
\usepackage{graphicx}
\usepackage{dcolumn}
\usepackage{bm}
\usepackage{cases}
\usepackage{color}
\usepackage{ulem}

\begin{document}

\title{Quantum particle confined to a thin-layer volume: Non-uniform convergence toward the curved surface}
\author{Yong-Long Wang$^{1,2,}$}
 \email{wangyonglong@lyu.edu.cn}
\author{Hong-Shi Zong$^{1,3,4,}$}
\email{zonghs@nju.edu.cn}
\address{$^{1}$ Department of Physics, Nanjing University, Nanjing 210093, China}
\address{$^{2}$ Department of Physics, School of Science, Linyi University, Linyi 276005, China}
\address{$^{3}$ Joint Center for Particle, Nuclear Physics and Cosmology, Nanjing 210093, China}
\address{$^{4}$ State Key Laboratory of Theoretical Physics, Institute of Theoretical Physics, CAS, Beijing 100190, China}

\date{\today}

\begin{abstract}
We clearly refine the fundamental framework of the thin-layer quantization procedure, and further develop the procedure by taking the proper terms of degree one in $q_3$ ($q_3$ denotes the curvilinear coordinate variable perpendicular to curved surface) back into the surface quantum equation. The well-known geometric potential and kinetic term are modified by the surface thickness. Applying the developed formalism to a toroidal system obtains the modification for the kinetic term and the modified geometric potential including the influence of the surface thickness.
\end{abstract}

\pacs{03.65.Ca, 02.40.-k, 68.65.-k}
\maketitle
\section{Introduction}
The thin-layer quantization formalism was first introduced in 1971 by H. Jensen and H. Koppe \cite{HJensen1971}, and generalized by R. C. T. da Costa (JKC) to investigate the quantum dynamics for a constrained single particle \cite{Costa1981} and for constrained multiple particles \cite{Costa1982}. In the three original papers, the fundamental framework of the JKC procedure was actually employed, but it was not explicitly defined.

With the development of the theoretical condensed matter physics, two dimensional (2D) curved systems are extensively investigated to study new physical effects that depend on both the curvature and the electromagnetic field, such as Aharonov-Bohm effect \cite{Bachtold1999, Noguchi2014, Silva2015}, quantum Hall effect \cite{Perfetto2007,Cresti2012}. Recently, some experiments were designed to investigate the geometric effects on the transport in photonic topological crystals \cite{Szameit2010}, on the proximity effects \cite{Kim2012} and on the electron states \cite{Onoe2012}. Both the theoretical and experimental developments have attracted tremendous interest in the generalization of the JKC procedure to discuss a curved system with an electromagnetic field \cite{Encinosa2006, Ferrari2008, BJensen2009, BJensen2010, Wang2014}. Under certain conditions, for the electromagnetic field a proper gauge should be chosen \cite{Ferrari2008}. At the same time, the presence of the electromagnetic field determines that the motion equation of the vector potential for the electromagnetic field should be included \cite{BJensen2009}. However, there is no an explicit fundamental framework of the JKC procedure to study the quantum equation, the chosen gauge and the motion of the vector potential simultaneously. The absence may lead to some calculational ambiguities \cite{Liu2007}. Generally, the curved system in an electromagnetic field can be described by a canonical action integral \cite{BJensen2009, Ortix2011}. By performing partial integration, the action can be divided into a volume integral and a surface integral. By varying the volume integral, the mentioned quantum equation can be obtained. In the general form, the absence of the fundamental framework of the JKC procedure maybe lead some calculational ambiguities for the simplifications of the integrals.

In the present paper, we will explicitly refine the fundamental framework of the JKC procedure. The procedure determines that the limit $q_3\to 0$ ($q_3$ is the curvilinear coordinate variable perpendicular to the curved surface) must be performed after calculating all derivatives with respect to $q_3$, and the limit $d\to 0$ ($d$ denotes the thickness of the curved surface) must be done after integrating all integrations of $q_3$. Employing the framework, we reconsider a spin-less charged particle confined on a curved surface in an electromagnetic field. For the considered system, the Coulomb gauge, which is chosen for the vector potential of the electromagnetic field, the motion of the vector potential, which couples to the three-dimensional (3D) electric currents $\vec{J}$, and the Schr\"{o}dinger equation are together originally defined in (3D) curved space. It is more physical and actual that a curved system has a certain thickness \cite{Diep2015}. We develop the fundamental framework by taking the suitable terms of degree one in $q_3$ back into the surface quantum dynamics. These terms modify the well-known geometric potential and kinetic term. These modifications can approximately describe the effects of the surface thickness. As an example for the applications of the developed procedure, we consider a spin-less charged particle constrained in a thin toroidal volume in the presence of an electromagnetic field.

This paper is organized as follows: in Sec. II, the fundamental framework of the JKC procedure is explicitly refined. In Sec. III, a spin-less charged particle bounded on a curved surface with an electromagnetic field is reconsidered in the refined framework. In Sec. IV, we develop the JKC formalism to primitively include the effects of the thickness of curved surface. In Sec. V, using the developed JKC procedure we investigate a spin-less charged particle confined in a thin toroidal volume with an electromagnetic field. In Sec. VI, we conclude and discuss the paper.
\section{The fundamental framework of the JKC procedure}
\begin{figure}[htbp]
\centering
\includegraphics[scale=0.37]{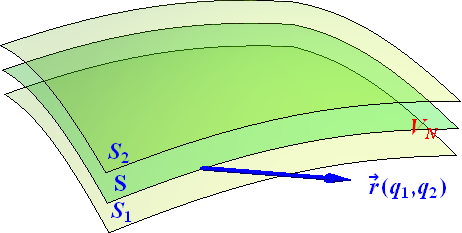}\\
\caption{\footnotesize (Color online) Schematics of the main surface $S$, the subspace $V_N$ and two auxiliary surfaces $S_1$ and $S_2$. $S$ is described by $\vec{r}(q_1,q_2)$. $V_N$ is enclosed by $S_1$ and $S_2$.}\label{Subspace}
\end{figure}

For the sake of convenient statement, we first define a subspace $V_N$ which is enclosed by two parallel surfaces $\mathcal{S}_1$ and $\mathcal{S}_2$ with a certain distance $d$. The main surface $\mathcal{S}$ has a unique distance $d/2$ to $\mathcal{S}_1$ and $\mathcal{S}_2$. They are sketched in Fig. \ref{Subspace}. $\mathcal{S}$ is parametrized by $\vec{r}(q_1,q_2)$. $V_N$ can be described by
\begin{equation}\label{Parametrization 1}
\vec{R}(q_1,q_2,q_3)=\vec{r}(q_1,q_2)+q_3\vec{n}(q_1,q_2),
\end{equation}
where $\vec{n}(q_1,q_2)$ is a unit vector perpendicular to $\mathcal{S}$.

In $V_N$, the metric tensor is defined by
\begin{equation}\label{Metric Tensor 1}
G_{ij}=
\left (
\begin{array}{ccc}
G_{11} & G_{12} & 0\\
G_{21} & G_{22} & 0\\
0 & 0 & 1\\
\end{array}
\right ), (i,j=1,2,3),
\end{equation}
where $G_{ij}=\frac{\partial\vec{R}}{\partial q^i}\cdot\frac{\partial\vec{R}}{\partial q^j}$. Squeezed on $\mathcal{S}$, the metric tensor \eqref{Metric Tensor 1} is simplified as
\begin{equation}\label{Metric Tensor 2}
\tilde{G}_{ij}=
\left (
\begin{array}{ccc}
g_{11} & g_{12} &0\\
g_{21} & g_{22} &0\\
0 & 0 &1
\end{array}
\right )=\lim_{q_3\to 0}(G_{ij}),
\end{equation}
where $g_{ab}=\frac{\partial\vec{r}}{\partial q^a}\cdot\frac{\partial\vec{r}}{\partial q^b}$, $(a, b=1,2)$. The relation between $G$ and $g$ is
\begin{equation}\label{Relation 1}
G=f^2g,
\end{equation}
where $G$ is the determinant of the matrix $G_{ij}$, $g$ is the determinant of the matrix $\tilde{G}_{ij}$, and the factor $f$ is
\begin{equation}\label{f}
f=1+2Mq_3+Kq_3^2,
\end{equation}
wherein $M$ is the mean curvature and $K$ is the Gaussian curvature, they are defined by
\begin{equation}\label{MK}
M=\frac{1}{2}\mathrm{Tr}(\alpha),\quad K=\det(\alpha).
\end{equation}
In Eq. \eqref{MK}, the elements of the Weingarten curvature matrix $\alpha$ are expressed as
\begin{equation}\label{Alpha}
\begin{split}
& \alpha_{11}=\frac{1}{g}(g_{12}h_{21}-g_{22}h_{11}), \alpha_{12}=\frac{1}{g}(g_{21}h_{11}-g_{11}h_{21}),\\
& \alpha_{21}=\frac{1}{g}(g_{12}h_{22}-g_{22}h_{12}), \alpha_{22}=\frac{1}{g}(g_{12}h_{21}-g_{11}h_{22}),
\end{split}
\end{equation}
where $h_{ab}$ are the coefficients of the second fundamental form, $h_{ab}=\vec{n}\cdot\frac{\partial^2\vec{r}}{\partial q^a\partial q^b}$, wherein $\vec{n}$ is the unit vector perpendicular to $\mathcal{S}$ \cite{Ono2009},
\begin{equation}\label{Normal}
\vec{n}=\frac{\frac{\partial\vec{r}}{\partial q^1}\times\frac{\partial\vec{r}}{\partial q^2}} {|\frac{\partial\vec{r}}{\partial q^1}\times\frac{\partial\vec{r}}{\partial q^2}|}.
\end{equation}

In the light of the three original papers \cite{HJensen1971, Costa1981, Costa1982}, it is straightforward to learn that the final aim of the JKC procedure is to squeeze the particle on $\mathcal{S}$, and to keep the effects induced by the surface curvature in the surface quantum dynamics as much as possible. According to the aim, we clearly refine the fundamental framework of the JKC procedure as follows:\\
(1) {\it A curved system (including dynamical equations, gauge conditions and so on) is originally defined in $V_N$.}\\
(2) {\it In terms of the metric tensor $G_{ij}$ defined in $V_N$, one calculates all the curvilinear coordinate derivatives those appear in Step (1).}\\
(3) {\it To perform the limit $q_3\to 0$ to remove all the terms depending on $q_3$ that present in Step (2), and accomplish to separate the dynamical equation into surface and normal components.}

The feasibility of Step (3) is ensured by the introduction of the squeezing potential \cite{HJensen1971}
\begin{equation}\label{Squeezing Potential}
V_{\lambda}(q_3)=
\begin{cases}
0,\quad q_3=0,\\
\infty,\quad q_3\neq 0,
\end{cases}
\end{equation}
which squeezes the particle on $\mathcal{S}$. In the fundamental procedure, it is rather obvious but very important that the limit $q_3\to 0$ must be performed after calculating all derivatives with respect to $q_3$, and the limit $d\to 0$ must be done after integrating all integrations of $q_3$. The two stipulations can preserve the effects induced by the surface curvature in the surface dynamics as much as possible, and can help user to avoid some calculational ambiguities.
\section{A spin-less charged particle confined on 2D curved surface in an electromagnetic field}
In the refined framework, we reconsider a spin-less charged particle confined on $\mathcal{S}$ in an electromagnetic field. The system can be described by the Schr\"{o}dinger equation that is
\begin{equation}\label{Schrodinger Equation 5}
i\hbar D_t\psi=-\frac{\hbar^2}{2m}D_iD^i\psi+V_{\lambda}(q_3)\psi,
\end{equation}
where $m$ is the mass of particle, $\psi$ is a wave function, $V_{\lambda}(q_3)$ is the potential \eqref{Squeezing Potential}, $D_t=\partial_t+\frac{ie}{\hbar}A_0$, and $D_i=\nabla_i+\frac{ie}{\hbar}A_i$, wherein $-e$ is the charge of particle. For the electromagnetic field, we choose the Coulomb gauge. In $V_N$, we expand the Schr\"{o}dinger equation \eqref{Schrodinger Equation 5} and the Coulomb gauge in the following forms
\begin{equation}\label{Schrodinger Equation 6}
i\hbar D_t\psi= -\frac{\hbar^2}{2m}[ \frac{1}{\sqrt{G}}\partial_i (\sqrt{G}G^{ij}\partial_j\psi)
+\frac{2ie}{\hbar}G^{ij}A_j \partial_i\psi-\frac{e^2}{\hbar^2}G^{ij}A_iA_j\psi]
+V_{\lambda}(q_3)\psi,
\end{equation}
and
\begin{equation}\label{Coulomb Gauge 1}
\vec{\nabla}\cdot\vec{A}=\frac{1}{\sqrt{G}} \partial_i(\sqrt{G}G^{ij}A_j) =0, (i,j=1,2,3),
\end{equation}
where $G^{ij}$ is the reciprocal of the matrix $G_{ij}$ defined by Eq. \eqref{Metric Tensor 1}. In order to decompose the quantum equation \eqref{Schrodinger Equation 6} into surface and normal components, we introduce a new wave function $\chi(q_1,q_2,q_3)=\chi_s(q_1,q_2)\chi_t(q_3)$. The conservation of the norm gives the relation:
\begin{equation}\label{PCR}
\psi=f^{-\frac{1}{2}}\chi,
\end{equation}
where $f$ is defined by Eq. \eqref{f}. Substituting Eq. \eqref{PCR} into Eq. \eqref{Schrodinger Equation 6}, and implementing the limit $q_3\to 0$, we can rewrite the Schr\"{o}dinger equation Eq. \eqref{Schrodinger Equation 6} as
\begin{equation}\label{Schrodinger Equation 7}
\begin{split}
i\hbar D_t\chi=& -\frac{\hbar^2}{2m}[\frac{1}{\sqrt{g}}\partial_a (\sqrt{g}g^{ab}\partial_b\chi)
+\frac{2ie}{\hbar}g^{ab}A_b\partial_a\chi
-\frac{e^2}{\hbar^2}g^{ab}A_aA_b\chi +\partial_3\partial^3\chi+ \frac{2ie}{\hbar}A^3\partial_3\chi\\
& -\frac{2ie}{\hbar}A^3M\chi -\frac{e^2}{\hbar^2}A_3A^3\chi]+V_g\chi+V_{\lambda}(q_3)\chi,
\end{split}
\end{equation}
where $(a,b=1,2)$ and $V_g$ is the well-known geometric potential \cite{HJensen1971} defined by
\begin{equation}\label{Geometric Potential}
V_g=-\frac{\hbar^2}{2m}[M^2-K],
\end{equation}
wherein $M$ is the mean curvature, $K$ is the Gaussian curvature in Eq. \eqref{MK}. By limiting $q_3\to 0$, the Coulomb gauge \eqref{Coulomb Gauge 1} can be rewritten as
\begin{equation}\label{Gauge 2}
\vec{\nabla}\cdot\vec{A}=\frac{1}{\sqrt{g}}\partial_a (\sqrt{g}g^{ab}A_b)+\partial_3A^3+2MA^3=0,
\end{equation}
where $(a,b=1,2)$ and $M$ is the mean curvature. It shows that the term $2MA^3$ in Eq. \eqref{Gauge 2} is given by
\begin{equation}\label{2MA}
\lim_{q_3\to 0}\frac{1}{\sqrt{G}}(\partial_3\sqrt{G})A^3=\lim_{q_3\to 0}\frac{1}{f}(\partial_3f)A^3=2MA^3.
\end{equation}
In contrast to the result in \cite{Ferrari2008}, the contribution of the Coulomb gauge keeps the term $-\frac{2ie}{\hbar}A^3M\chi$ in the left hand side of Eq. \eqref{Schrodinger Equation 7}, and vanishes the terms $\frac{ie}{\hbar}\frac{1}{\sqrt{g}} \partial_a(\sqrt{g}g^{ab}A_b)\chi$ and $\frac{ie}{\hbar}(\partial_3A^3)\chi$.

In stationary situation, the motion of the vector potential $\vec{A}$ in the Coulomb gauge is
\begin{equation}\label{Maxwell Equation}
\nabla^2\vec{A}=[\frac{1}{\sqrt{G}}\partial_i (\sqrt{G}G^{ij}\partial_j)]\vec{A}=-\mu\vec{J},
\end{equation}
where $(i,j=1,2,3)$, $\mu$ is the permeability of the material, $\vec{J}$ are 3D electric currents. The vanishing of the $J_3$ component current can supply an equation:
\begin{equation}\label{J301}
\frac{1}{\sqrt{G}}\partial_a(\sqrt{G}G^{ab}\partial_b)A_3 +\frac{1}{\sqrt{G}}\partial_3(\sqrt{G}\partial^3)A_3=0.
\end{equation}
Squeezing on $\mathcal{S}$, we can rewrite Eq. \eqref{J301} as
\begin{equation}\label{J302}
\frac{1}{\sqrt{g}}\partial_a(\sqrt{g}g^{ab}\partial_b)A_3 +\partial_3\partial^3A_3+2M\partial^3A_3=0
\end{equation}
with $(a,b=1,2)$. From Eq. \eqref{Schrodinger Equation 7}, it is straightforward that the vanishing of the terms of $A^3$ is necessary and sufficient to decouple the quantum dynamics \eqref{Schrodinger Equation 7} into a surface dynamics and a normal dynamics. Fortunately, it is apparently possible to find a restricted gauge function $f(q_1,q_2,q_3)$, which satisfies the gauge transformation $A^{\prime 3}(q_1,q_2,q_3)=A^3(q_1,q_2,q_3) +\partial_3f(q_1,q_2,q_3)=0$, in the Coulomb gauge. The Coulomb gauge implies the identity $\partial_3\partial^3f=0$. Furthermore, we have $\partial_3A^{\prime 3}=0$ and $\partial_3A^3=0$. These transformed results satisfy Eq. \eqref{J302}. As $J_3\neq 0$, the Coulomb gauge obviously determines that we can not find a restricted gauge function to satisfy $A^{\prime 3}=0$ and
\begin{equation}\label{J3N0}
\frac{1}{\sqrt{g}}\partial_a(\sqrt{g}g^{ab}\partial_b)A^{\prime}_3 +\partial_3\partial^3A^{\prime}_3+2M\partial^3A^{\prime}_3\neq 0
\end{equation}
with $(a,b=1,2)$, simultaneously. In consequence, the vanishing of the $J_3$ component current is necessary to decompose the Schr\"{o}dinger equation \eqref{Schrodinger Equation 7} into a surface equation
\begin{equation}\label{Schrodinger Equation 7-1}
i\hbar\partial_t\chi_s=-\frac{\hbar^2}{2m}[\frac{1}{\sqrt{g}}\partial_a(\sqrt{g} g^{ab}\partial_b\chi_s)
+\frac{2ie}{\hbar}g^{ab}A_b \partial_a\chi_s-\frac{e^2}{\hbar^2}g^{ab}A_aA_b\chi_s]
+V_g\chi_s+eA_0\chi_s,
\end{equation}
and a normal equation
\begin{equation}\label{Schrodinger Equation 7-2}
i\hbar\partial_t\chi_t =-\frac{\hbar^2}{2m}\partial_3\partial^3\chi_t +V_{\lambda}(q_3)\chi_t.
\end{equation}
It is noteworthy that the Schr\"{o}dinger equation \eqref{Schrodinger Equation 7}, the Coulomb gauge \eqref{Gauge 2}, and the motion equation of the vector potential \eqref{Maxwell Equation} together describe the spin-less charged particle confined on $\mathcal{S}$ in the presence of an electromagnetic field including its sources $\vec{J}$. At the same vein, they together determine that the vanishing of the $J_3$ component current is necessary to the validity of the JKC procedure. And they are originally defined in $V_N$, we can not simplify them by prematurely limiting $q_3\to 0$.

In a general form, the previously discussed system can be described by a canonical action integral \cite{Ferrari2008, BJensen2009, Ortix2011} as
\begin{equation}\label{Action 3}
S=\int_{V_N}[-i\hbar\psi^{*}D_t\psi
+\frac{\hbar^2}{2m}(\vec{D}\psi)^{*} \cdot(\vec{D}\psi)+V_{\lambda}(q_3)|\psi|^2],
\end{equation}
where $D_t=\partial_t+\frac{ie}{\hbar}A_0$, $\vec{D}=\vec{\nabla}+\frac{ie}{\hbar}\vec{A}$, $\psi$ is a wave function and $V_{\lambda}(q_3)$ is the squeezing potential \eqref{Squeezing Potential}. By performing partial integration, in the Coulomb gauge the action \eqref{Action 3} can be divided into a volume integral
\begin{equation}\label{Action 3-1}
S_v=\int_{V_N}\{-i\hbar\psi^{*}D_t\psi
-\frac{\hbar^2}{2m}\psi^*[\nabla_i\nabla^i\psi -\frac{e^2}{\hbar^2}\vec{A}^2\psi +\frac{2ie}{\hbar}A^i\nabla_i\psi +V_{\lambda}(q_3)\psi]\},
\end{equation}
and a closed surface integral
\begin{equation}\label{Action 3-2}
S_s=\frac{\hbar^2}{2m}\oint_{\partial V_N}(\psi^{*}\vec{D}\psi).
\end{equation}
Varying the action $S_v$ in \eqref{Action 3-1}, we obtain the Schr\"{o}dinger equation for $\psi$ in the Coulomb gauge as Eq. \eqref{Schrodinger Equation 6}. It is easy to decompose into the surface dynamics \eqref{Schrodinger Equation 7-1} and the normal dynamics \eqref{Schrodinger Equation 7-2} by repeating the above adopted procedure. The surface integral \eqref{Action 3-2} plays the role of a boundary condition, which vanishes, and that the Schr\"{o}dinger equation \eqref{Schrodinger Equation 6} is stipulated only by the volume integral \eqref{Action 3-1}. With the limit $d\to 0$, the vanishing of the integral \eqref{Action 3-2} is equivalent to
\begin{equation}\label{Condition 3}
\lim_{d\to 0}[\psi^*(\partial_3-\frac{ie}{\hbar}A_3)\psi]_{\frac{d}{2}} =\lim_{d\to 0}[\psi^*(\partial_3-\frac{ie}{\hbar}A_3)\psi]_{-\frac{d}{2}}.
\end{equation}
It is easy to see that the vanishing of the integral \eqref{Action 3-2} is trivially satisfied by the smoothness of $\psi$ and $\partial_3\psi$ passing through $\mathcal{S}$ \cite{Ortix2011}. As a conclusion, the JKC method can be used with the continuity of $\psi$ and $\partial_3\psi$ passing through $\mathcal{S}$. Under the conditions, an arbitrary boundary condition, which is imposed on the normal fluctuation of the wave function, does not endanger the validity of the JKC approach.

If the limit $d\to 0$ is brought into the surface integral and replaced by $q_3\to 0$, we would obtain a trivial identity
\begin{equation}\label{Condition 4}
\lim_{q_3\to 0}[\psi^*(\partial_3-\frac{ie}{\hbar}A_3)\psi] =\lim_{q_3\to 0}[\psi^*(\partial_3-\frac{ie}{\hbar}A_3)\psi],
\end{equation}
or obtain a boundary condition
\begin{equation}\label{Condition 5}
\lim_{q_3\to 0}[\psi^*(\partial_3-\frac{ie}{\hbar}A_3)\psi]=0,
\end{equation}
the later condition is very more strict than the condition Eq. \eqref{Condition 3}. On physical grounds one can argue that it is more physically consistent to assume certain variation $\delta\psi$ on $\partial V_N$. In the case, the artificially enhanced condition gives
\begin{equation}\label{Condition 6}
\lim_{q_3\to 0}(\partial_3-\frac{ie}{\hbar}A_3)\psi=0,
\end{equation}
which means that an imposed Nuemann type boundary condition will invalidate the JKC procedure except where the curved surface has a constant mean curvature \cite{BJensen2009}. In order to escape some unnecessary ambiguities, the limit $d\to 0$ should be performed after calculating all curvilinear coordinate integrals, especially the integrations of $q_3$. By following the stipulation, it is easy to check that the surface integral \eqref{Action 3-2} can not contribute a volume integral term $-\frac{ie\hbar}{m}A^3M\chi^*\chi$ to the volume integral \eqref{Action 3-1}. The convenient contribution stems from that the limit $d\to 0$ is brought into the integral and replaced by $q_3\to 0$ performed prematurely.
\section{The modification induced by the surface thickness}
It is the central result of the JKC procedure that the geometric potential induced by the surface curvature appears in the surface dynamics. The process is probably severe and breaks with natural limits set by the uncertainty principle, but it has been demonstrated that the attractive geometric potential is valid and important to some curved systems \cite{Koshino2005, Fujita2005, Shima2009}. On actual grounds, the curved surface with certain thickness is the real existence. To the best of our knowledge, it is still not completely clear how to introduce the modifications induced by the surface thickness into the surface dynamics. In keeping within the scope of the fundamental procedure refined in Sec. II, we further extend the JKC framework by adding a step that is\\
(4){\it { To take the certain terms (they were vanished in Step (3)) of degree one in $q_3$ back into the surface quantum dynamics obtained in Step (3).}} It is necessary to indicate that here the relived terms are not all the first degree terms with respect to $q_3$. The selected basis is that each of the revived terms must have counterpart terms appearing in the surface Schr\"{o}dinger equation in Step 3, and it with its counterparts are together deduced from a same original term, the original term can be found in the original Schr\"{o}dinger equation in Step 1.

Performing Step (4), we extend the surface Schr\"{o}dinger equation \eqref{Schrodinger Equation 7-1} as
\begin{equation}\label{Thickness 1}
i\hbar\partial_t\chi_s= -\frac{\hbar^2}{2m}[\frac{1}{\sqrt{g}} \partial_a (\sqrt{g}g^{ab}\partial_b\chi_s)
+\frac{2ie}{\hbar} g^{ab}A_b(\partial_a\chi_s) -\frac{e^2}{\hbar^2}g^{ab}A_aA_b\chi_s]
+H^{\prime}\chi_s
+V_g^{\prime}\chi_s+eA_0\chi_s.
\end{equation}
Here $V_g^{\prime}$ is a new geometric potential, which is the well-known geometric potential extended to include the modification of the thickness of the curved surface, in the following form
\begin{equation}\label{Thickness GP}
V_g^{\prime}=V_g(1-4Mq_3),
\end{equation}
where $V_g$ is the well-known potential \eqref{Geometric Potential}, $M$ is the mean curvature. In Eq. \eqref{Thickness 1}, the contribution of degree one in $q_3$ to the kinetic term, $H^{\prime}$ is calculated as
\begin{equation}\label{Modification Kinetic}
H^{\prime}=\frac{\hbar^2}{2m}q_3\{\frac{1}{\sqrt{g}}\partial_a [\sqrt{g}g^{ab}(\partial_bM)-\sqrt{g}w^{ab}\partial_b]
+\frac{2ie}{\hbar}A_a[g^{ab}(\partial_bM)-w^{ab}\partial_b] +\frac{e^2}{\hbar^2}w^{ab}A_aA_b
-2(\partial_aM)g^{ab}\partial_b\}
\end{equation}
with $w^{ab}=g^{\prime ab}-Mg^{ab}$, where $g^{\prime ab}$ is determined by the expression $G^{ab}\approx g^{ab}+g^{\prime ab}q_3+\cdots$, wherein $G^{ab}$ is the reciprocal of $G_{ab}$, and $g^{ab}$ is the inverse of $g_{ab}$. It is worthwhile to note that $q_3$ in Eqs. \eqref{Thickness GP} and \eqref{Modification Kinetic} is a relative infinitesimal constant rather than a variable. This conclusion is determined by the limitation of the extended JKC procedure. Obviously, as $q_3=0$ the modified geometric potential $V_g^{\prime}$ becomes the well-known geometric potential $V_g$, the modification for the kinetic term vanishes, and Eq. \eqref{Thickness 1} is the same as in \cite{Ferrari2008}. When the electromagnetic field is disappeared, $A_{a}=0$, Eq. \eqref{Thickness 1} is different from the counterpart originally given in \cite{Costa1981} with the modification induced by the surface thickness. The modifications is rather obvious to $V_g$ under certain conditions as shown in Fig. \ref{TMGP} (a). In the limited case $q_3=0$, the equation \eqref{Thickness 1} is completely equivalent to the result in \cite{Costa1981} when the electromagnetic field is vanished. The presence of the geometrical potential newly defines the geometric momentum \cite{Liu2011} to replace the usual momentum. It is interesting to further study the modification induced by the thickness of surface to the geometric momentum.
\section{A spin-less charged particle confined in a thin toroidal volume with an electromagnetic field}
\begin{figure}[htbp]
  \centering
  \includegraphics[scale=0.33]{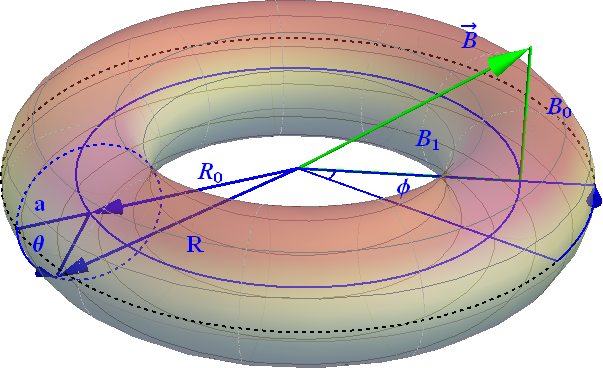}\\
  \caption{\footnotesize (Color online) A torus with a minor radius $a$ and a major radius $R_0$ in a constant electromagnetic field $\vec{B}$. The magnetic field $\vec{B}$ can be separated into $\vec{B}_1$, which lies in the plane determined by $R_0$ and $\phi$, and $\vec{B}_0$, which is normal to the circle defined by $R_0$ and $\phi$.}\label{Torus}
\end{figure}

A torus is a mathematical topological geometry. The particular topology is a test bed for models on curved surfaces \cite{Encinosa2003, Encinosa2006}. In the curvilinear coordinate system $(\theta,\phi,q_3)$, a torus is put in an arbitrary constant magnetic field $\vec{B}$ described in Fig. \ref{Torus}.

Now let us to apply the developed JKC procedure to the toroidal system. Points on the toroidal surface can be parametrized as
\begin{equation}\label{Toroidal Surface}
\vec{r}(\theta,\phi)=W\vec{e}_\rho+a\sin\theta\vec{e}_z,
\end{equation}
and then points near the surface may be parametrized as
\begin{equation}\label{TorusE}
\vec{R}(\theta,\phi,q)=W\vec{e}_\rho +a\sin\theta\vec{e}_z+q_3\vec{e}_n,
\end{equation}
where $W=R_0+a\cos\theta$, $\vec{e}_n$ is the unit vector perpendicular to the surface, $q_3$ is the curvilinear coordinate variable normal to the surface. According to the definitions $G_{ab}=\partial_a\vec{R}\cdot\partial_b\vec{R}$, and $g_{ab}=\partial_a\vec{r}\cdot\partial_b\vec{r}$ $(a,b=1,2)$, we obtain
\begin{equation}\label{TSMetric}
g_{ab}=
\left (
\begin{array}{cc}
a^2 & 0\\
0 & W^2
\end{array}
\right ),
\end{equation}
and
\begin{equation}\label{TMetric}
G_{ab}=
\left (
\begin{array}{cc}
(a+q_3)^2 & 0\\
0 & (W+q_3\cos\theta)^2
\end{array}
\right ).
\end{equation}
The inverse matrices with respect to $G_{ab}$ and $g_{ab}$ are
\begin{equation}\label{ITSMetric}
g^{ab}=
\left (
\begin{array}{cc}
\frac{1}{a^2} & 0\\
0 & \frac{1}{W^2}
\end{array}
\right ),
\end{equation}
and
\begin{equation}\label{ITMetric}
G^{ab}=
\left (
\begin{array}{cc}
\frac{1}{(a+q_3)^2} & 0\\
0 & \frac{1}{(W+q_3\cos\theta)^2}
\end{array}
\right ),
\end{equation}
respectively. In contrast to $a$ and $R_0$, $q_3$ is relative infinitesimal, which can be interpreted that the toroidal surface is enough thin, we can approximate the matrix $G^{ab}$ as
\begin{equation}\label{ITMetric1}
G^{ab}\approx g^{ab}+g^{\prime ab}q_3,
\end{equation}
reserved only the term of degree one in $q_3$, where
\begin{equation}\label{PITMetric}
g^{\prime ab}=-2
\left (
\begin{array}{cc}
\frac{1}{a^3} & 0\\
0 & \frac{\cos\theta}{W^3}
\end{array}
\right ).
\end{equation}
From Eqs. \eqref{TSMetric}, \eqref{TMetric} and \eqref{Relation 1}, we can obtain that the factor $f$ in Eq. \eqref{f} is
\begin{equation}\label{Tf}
f=1+\frac{W+a\cos\theta}{aW}q_3 +\frac{\cos\theta}{aW}q_3^2,
\end{equation}
the mean curvature and the Gaussian curvature are
\begin{equation}\label{TMK}
M=\frac{W+a\cos\theta}{2aW},\quad K=\frac{\cos\theta}{aW},
\end{equation}
respectively.

In the Coulomb gauge, we can choose the components of the vector potential for the toroidal system as
\begin{equation}\label{TVectorP}
A_{\theta}=B_1a^2\sin\phi,\quad
A_{\phi}=B_0W^2-B_1aW\sin\theta\cos\phi,\quad
 A_q=0.
\end{equation}
The quantum equation can deduced from Eq. \eqref{Thickness 1} as
\begin{equation}\label{MTSSE}
\begin{split}
i\hbar\partial_t\chi_s=-\frac{\hbar^2}{2m}\{&\frac{1}{a^2} \partial_{\theta}^2\chi_s-\frac{\sin\theta}{aW} \partial_{\theta}\chi_s+\frac{1}{W^2} \partial_{\phi}^2\chi_s+\frac{2ie}{\hbar}B_1\sin\phi \partial_{\theta}\chi_s+\frac{2ie}{\hbar}(B_0-B_1 \frac{a}{W}\sin\theta\cos\phi)\partial_{\phi}\chi_s\\
& -\frac{e^2}{\hbar^2}[(B_1a\sin\phi)^2 +(B_0W-B_1a\sin\theta\cos\phi)^2]\chi_s\}+H^{\prime}\chi_s +V_g^{\prime}\chi_s.
\end{split}
\end{equation}
The salient feature of Eq. \eqref{MTSSE} is the presence of the modification induced by the surface thickness $q_3$ for the kinetic term, which is
\begin{equation}\label{MTKT}
\begin{split}
H^{\prime}=\frac{\hbar^2}{2m}q_3[&\frac{5W+a\cos\theta} {2a^3W}\partial_{\theta}^2+\frac{W+5a\cos\theta}{2aW^3} \partial_{\phi}^2+(\frac{R_0\sin\theta}{2a^2W^2} -\frac{3\sin\theta}{a^2W})\partial_{\theta} +\frac{ie}{\hbar}\frac{5W+a\cos\theta} {aW}B_1\sin\phi\partial_{\theta} \\ &+\frac{ie}{\hbar}\frac{W+5a\cos\theta}{aW^2} (B_0W-B_1a\sin\theta\cos\phi) \partial_{\phi} -\frac{R_0(a+R_0\cos\theta)}{2a^2W^3}-\frac{ie}{\hbar} \frac{R_0\sin\theta}{W^2}B_1\sin\phi\\
&-\frac{e^2}{\hbar^2}\frac{5W+a\cos\theta}{2W} B_1^2a\sin^2\phi-\frac{e^2}{\hbar^2}\frac{W+5a\cos\theta} {2aW}(B_0W-B_1a\sin\theta\cos\phi)^2],
\end{split}
\end{equation}
\begin{figure}[htbp]
  \centering
  \includegraphics[scale=0.433]{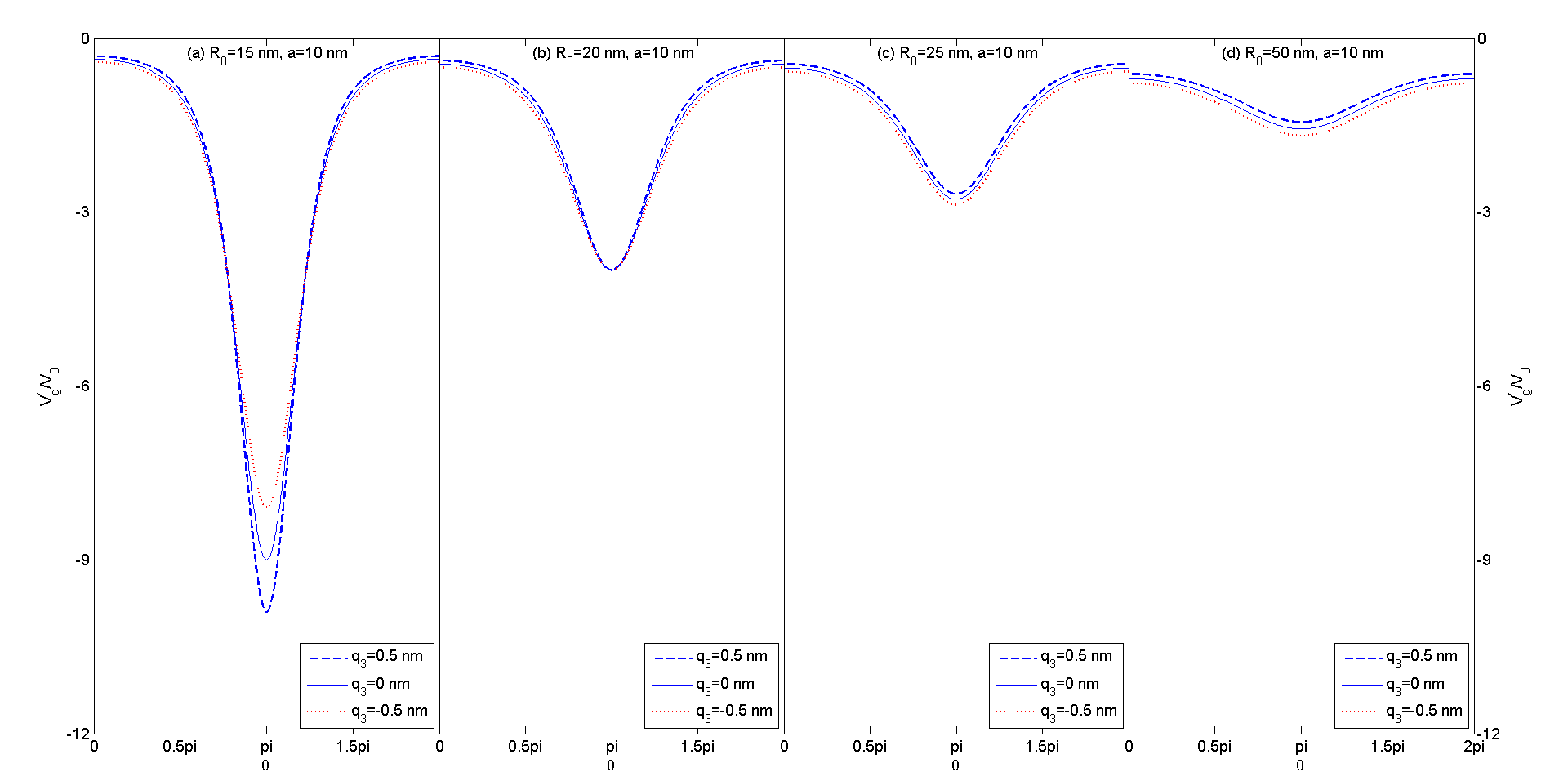}\\
  \caption{\footnotesize (Color online) Spatial profiles of $V_g^{\prime}$ in units of $V_0=\frac{\hbar^2}{8ma^2}$ with $q_3=0.5nm$, $0 nm$, $-0.5 nm$ at (a) $R_0=15 nm$, $a=10 nm$, (b) $R_0=20 nm$, $a=10 nm$, (c) $R_0=25 nm$, $a=10 nm$ and (d) $R_0=50 nm$, $a=10 nm$.}\label{TMGP}
\end{figure}
and the geometric potential is added by a first degree term of $q_3$ as below
\begin{equation}\label{MTGP}
V_g^{\prime}=-\frac{\hbar^2 R_0^2}{8ma^2W^2} +\frac{\hbar^2R_0^2(W+a\cos\theta)}{4ma^3W^3}q_3.
\end{equation}
Here the first term in the right hand side is the well-known geometric potential $V_g$ from the nonzero surface curvature, which is the same as in \cite{Ferrari2008}, the second term is the contribution of the first degree term of $q_3$. In other words, the second term in Eq. \eqref{MTGP} can be used to approximately describe the modification induced by the surface thickness to $V_g$. In order to visualize the modification induced by the surface thickness to $V_g$, the spatial profiles of the modified geometric potential $V_g^{\prime}$ for $q_3=0.5 nm$, $0 nm$, $0.5 nm$ in units of $\frac{\hbar^2}{8ma^2}$ are sketched in Fig. \ref{TMGP} at (a) $R_0=15 nm$, $a=10 nm$, (b) $R_0=20 nm$, $a=10 nm$, (c) $R_0=25 nm$, $a=10 nm$ and (d) $R_0=50 nm$, $a=10 nm$. When $q_3=0 nm$, $V_g^{\prime}$ trivially becomes equivalent to $V_g$. As shown in Fig.\ref{TMGP}, the downward peaks of $V_g^{\prime}$ and $V_g$ at $\theta=\pi$ decrease with increasing the major radius $R_0$ when the minor radius $a$ is fixed at a certain value. The relation between $a$ and $V_g^{\prime}$ or $V_g$ is implied in the unit $V_0=\frac{\hbar^2}{8ma^2}$. Fig.\ref{TMGP} (a) shows that the downward peaks of $V_g^{\prime}$ with $R_0=1.5a$ grow with increasing the surface thickness. But the surface thickness does not influence the amplitude of the downward peaks of $V_g^{\prime}$ with $R_0=2a$. It is described in Fig.\ref{TMGP} (c) and (d) that the contribution of the surface thickness is not significant to $V_g^{\prime}$. In the later three cases, one can use $V_g$ to replace $V_g^{\prime}$ completely under some certain conditions.

\section{Conclusions and discussions}
\indent We have explicitly refined the fundamental framework for the thin-layer quantization procedure that consists of three stages: (1) originally define the curved dynamics in the 3D subspace $V_N$, (2) subsequently calculate all various curvilinear coordinate derivatives appearing in the dynamics, (3) finally perform the limit $q_3\to 0$ to squeeze the particle on the curved surface and obtain the effective surface dynamics. Essentially, the fundamental framework is determined by the final aim of the JKC formalism. The aim is to squeeze the particle on the curved surface, simultaneously to keep the effects of the surface curvature in the expectant surface quantum equation as much as possible. Preserving the effects induced by the surface curvature is of essence to the JKC method, which naturally defines that the limit $q_3\to 0$ must be performed after calculating all derivatives with respect to $q_3$ in the dynamical equation, in the necessary gauge, and so on. In a general form, the limit $d\to 0$ must be done after integrating all integrations of $q_3$ in the canonical action integral and its deuterogenic integrals. The refined framework and the two stipulations can help user to avoid some ambiguities in the JKC procedure.

Using the refined fundamental framework, we have reconsidered a spin-less charged particle bounded on the curved surface in an electromagnetic field \cite{Ferrari2008}. The Coulomb gauge chosen for the electromagnetic field, the motion of the electromagnetic field and the Schr\"{o}dinger equation are originally defined in the 3D subspace $V_N$. These definitions determine that the vanishing of the $J_3$ component current is necessary to accomplish the decoupling of the electromagnetic field from the surface curvature and the separability of the Shcr\"{o}dinger equation into surface and normal dynamics. In the general form of the canonical action integral, the stipulation on the limit $d\to 0$ can guarantee that an arbitrary boundary condition, which is imposed on the normal fluctuation of the wave function, does not influence the validity of the JKC formalism provided that the wave function and its derivative with respect to $q_3$ smoothly pass through the curved surface \cite{Ortix2011}.

Furthermore, we have primitively considered the effects of the thickness of the curved surface by adding a step to the the fundamental framework. The step is to take the proper terms of degree one in $q_3$ back into the expectant surface quantum equation. These terms modify the well-known geometric potential and the kinetic term to include the effects of the surface thickness. Using the developed JKC procedure, we have investigated a spin-less particle squeezed in a thin toroidal volume, and have obtained the modification for the kinetic term and the modified geometric potential including the effects of the surface thickness. It shows that the surface thickness substantially affects the geometric potential under some special conditions.

\section{Acknowledgment}
\indent We thank U. Jentschura very much for very helpful suggestions. This work is supported by the National Natural Science Foundation of China (under Grant No. 11047020, No. 11404157, No. 11274166, No. 11275097, No. 11475085, and No. 11535005), and the Natural Science Foundation of Shandong Province of China (under Grant No. ZR2012AM022, and No. ZR2011AM019).

\end{document}